# Non-additive entropy: Reason and conclusions

**Miriam Lemanska**

Soreq NRC, Yavne 81800, Israel

E-Mail: lemanska@bezeqint.net

______________________________________________________________________

**Abstract:** In this work the non-additive entropy is examined. It appears in isolated particle systems composed of few components. Therefore, the mixing of isolated particle systems $S=S_1+S_2$ has been studied. Two cases are considered $T_1=T_2$ and $T_1 \ne T_2$, where $T_1, T_2$ are the initial temperatures of the system $S_1$ and $S_2$ respectively. The concept of similar systems containing interacting particles is introduced. These systems are defined by a common temperature and an identical time evolution process, i.e. the approach to the same thermodynamic equilibrium.

The main results are:

1) The properties of the similar particle systems yield the non-additive entropy and free energy. The Gibbs Paradox is not a paradox.
2) The relation between the initial temperatures $T_1$ and $T_2$ governs the mixing process.
3) In the two cases $T_1=T_2$, $T_1 \ne T_2$ mixing of the systems $S_1$, $S_2$ results in a uniform union system $S=S_1+S_2$. The systems $S$, $S_1$, $S_2$ are similar one to the other.
4) The mixing process is independent of the extensive quantities (volume, particle number, energy) and of the particle type. Only the mean energy plays an important role in the mixing of the systems $S_1$, $S_2$.
5) Mixing in the case $T_1 \ne T_2$ is in essence a thermalization process, but mixing in the case $T_1=T_2$ is not a thermodynamic process.
6) Mixing is an irreversible process.

______________________________________________________________________

## 1. Introduction

During the last years some papers have been published in which again Gibbs Paradox (non-additive entropy), the concept of entropy and the mixing process of particle systems were considered, see for example [1-6]. From the definition of additivity (or non-additivity) follows that this notion is related to a system composed of several components, $S = \sum_i S_i$ $i \ge 2$. $S_i$ are isolated systems containing interacting particles.

The union system S is a mixture of the $S_i$ ones. This linkage plays a fundamental role in our consideration. Let $S_i$ contain particles of different types, then $S=\cup_i S_i$ presents an isolated system containing interacting particles of different types. The aim of this work is to find the reason for the non-additive entropy via the study of the properties of a mixture S. To simplify our consideration the mixture $S=S_1+S_2$ will be examined. Two cases are here studied: (i) $T_1=T_2$, (ii) $T_1 \neq T_2$ ($T_1$, $T_2$ are the temperatures of $S_1$, $S_2$). Mixing of the systems $S_1$ and $S_2$ yields the system $S=S_1+S_2$ defined in Sec. 2.1.

Note that the systems S, $S_1$, $S_2$ in the case (i) are in thermal equilibrium, i.e. the mixture system S has the temperature $T=T_1=T_2$. We ask what is the relation between the particle distributions over the energy levels ($0 \leq e_i \leq 1$ i=1,2,...I) in the whole system S and in its components $S_1$, $S_2$.

The case (ii) presents a thermalization process. The question of interest is how the changes in the particle distribution over the energy levels lead to a thermal equilibrium (equality $T_1=T_2$ at later times).

In order to understand the results of the mixing process in the cases (i) and (ii) the concept of similar systems is introduced in Sec. 2.2. Properties of similar systems are given by Theorems 1 and 2. With the help of Theorem 2 we have obtained a solution of the mixing problem for the case (i), $T_1=T_2$. This solution allows to understand the thermal equilibrium.

The case (i) is studied in Sec. 2.2 and (ii) in Sec. 2.3. To achieve our purpose the particle distribution and the system state are expressed in terms of the occupation probabilities, Eqs. (2.5) and (2.6).

In Secs. 2.2-2.4 the following statements are proven:

1) Mixing in the cases (i) and (ii) forms an isolated, uniform union system, $S=S_1+S_2$, which is similar to its components $S_1$, $S_2$.

2) In the case (ii) the uniform union system S is obtained simultaneously with the attainment of the thermal equilibrium ($T_1=T_2$).

3) The entropy is a non-additive thermodynamic quantity. This non-additivity results from the properties of similar particle systems.

4) Mixing is an irreversible process.

These results are discussed in Sec. 3.

In the Appendix some results concerning the approach to the thermodynamic equilibrium are recalled from [7].

## 2. Mixing of two isolated systems containing interacting particles

### 2.1 Statement of the problem

Let there be given two systems $S_1$, $S_2$ isolated in the "universum" U and consisting of the interacting particles (see Fig. 1).

The quantities associated with the systems $S_1$ and $S_2$ are denoted by:

$T_k$ - the temperature, $E_k$ - the energy, $V_k$ - the volume, $N_k$ - the number of particles, k=1,2.

Let us remove the partition wall XY (Fig. 1). Then the obtained system S is isolated in U and is defined as follows:

(2.1)    $S = S_1 + S_2$

(2.2)    $V = V_1 + V_2$

(2.3)    $E = E_1 + E_2$

(2.4)    $N = N_1 + N_2$

In Secs. 2.2 and 2.3 we examine the question how the mixing of systems $S_1$ and $S_2$ forms a homogenous system S.

In other words, we should find the conditions under which the particle distributions of the components $S_1$ and $S_2$ are the same.

Because in the sequel the particle distribution will be expressed by the occupation probabilities, we give their definitions. Let $0 \leq e_i \leq 1$ i=1,...,I be the basic energy levels and $n_{k,i}$ k=1,2 the number of particles occupying the energy level $e_i$. The occupation probabilities are:

(2.5)    $p_{k,i} = n_{k,i}/N_k$  i=1,2,...,I , k=1,2

The state of a system is defined by the finite sequence

(2.6)    $p = (p_1 p_2, ... p_I)$

The mean energy $E_k^* = E_k/N_k$ k=1,2 expressed in the terms (2.5) is

(2.7)    $E_k^* = \sum_i^I e_i p_{k,i}$  k=1,2

We assume, in Secs. 2.2 and 2.3 the considered systems have the same energy level structure.

## 2.2. Mixing of two isolated particle systems having the same temperature, $T_1=T_2$.

**Similar systems of interacting particles.**

First we deal with the concept of similar particle systems, which is necessary for understanding the mixing problem of two isolated particle systems.

The relations between the system S and its components $S_1,S_2$ being in thermal equilibrium, see Sec. 2.3, suggest the concept of similar systems of interacting particles.

Definition 1. The isolated particle systems $\{S_l\}$ l=1,2,…L,L≥2, being in non-equilibrium state are said to be similar (one to the other) if they have common temperature T (= common mean energy $E^*$) and the time evolution process is the same.

By the same (identical) evolution processes (approach to the thermodynamic equilibrium) we understand that the systems of interest pass through the same states $p(t)=(p_1(t),…,p_l(t))$ for $t_0 \leq t < \infty$ during the approach to the same thermodynamic equilibrium (= the same Boltzmann distribution). The time evolution process has been studied in [7]. Some results are recalled in Appendix.

Remark 1. The case of isolated systems having common temperature and being in thermodynamic equilibrium is trivial. Of course these systems are similar. The common temperature determines the common Boltzmann distribution at any time t.

Let the collection of isolated systems $\{S_l\}$ l=1,2,…L,L≥2 be in non-equilibrium state.

Theorem 1. Systems $\{S_l\}$ are similar one to the other iff they have a common temperature T.

*Proof*. Necessary condition. Assume that the $\{S_l\}$ systems are similar, then from Definition 1 follows that they have a common temperature T.

Sufficient condition. We assume that the systems $\{S_l\}$ have a common temperature T (= common mean energy $E^*$). We should show that the systems $\{S_l\}$ have the same time evolution process.

The proof is based on the results of [7], some of which related to our problem are quoted in the items of the Appendix. Let $t_0$ be the initial time of the evolution process occuring in the systems belonging to the class $\{S_l\}$. To achieve our purpose we shall consider the consequences of the above assumptions: The systems $\{S_l\}$ are isolated and have the common temperature T.

Isolation property. This assumption allows to apply Eq. (A.2) to the investigation of the time evolution process occuring in the systems $\{S_l\}$.

Of course the statements included in item (1) are common to systems $\{S_l\}$.

Common temperature. The mean energy $E^*$ is common to the systems $\{S_l\}$. We know from item (1) that the solutions p(t) of Eq. (A.2), $t_0 \leq t < \infty$, present the system states at the time t.

$p(t)=(p_1(t),...,p_I(t))$ establishes a point in I-dimensional Euclidean space, whose coordinates $p_i(t)$, i=1,...I, are the occupation probabilities at the time t. Those coordinates obey the energy conservation law for the systems $\{S_l\}$ with the same mean energy $E^*$.

In virtue of items (1) and (2) we affirm what follows.

The set P containing the solutions of Eqs. (A.2) is common to all systems of the class $\{S_l\}$ and establishes a common interval belonging to the same hyperline, the intersection of two hyperplanes $\sum_i p_i(t)=1$ and $\sum_i e_i p_i(t)=E^*$, $t_0 \leq t < \infty$.

Other important fact caused by the common temperature T is that the systems of the class $\{S_l\}$ approach to the same thermodynamic equilibrium, because the Boltzmann distribution is determined by the quantity $\beta=1/T$.

Items (3) and (4) completely determine the set P of the solutions p(t) of Eqs. (A.2).

P constitutes the interval $<p(t_0),p_B)$ which is closed under the thermodynamic equilibrium state. It is shown in [7] that the distribution $p_B$ is the fixed point of Eqs. (A.2) for the equilibrium state. The time evolution process in each system belonging to class $\{S_l\}$ takes place on the interval $<p(t_0),p_B>$, which forms the domain and the range of Eqs. (A.2).

D(A.2)=R(A.2) is common to the systems of the class $\{S_l\}$. Thus Theorem 1 is proven.

Note, Definition 1 and the proof of Theorem 1 are irrespective of the system volume, the particle number and the particle type. Only the common temperature (=the ratio E/N) plays an essential role in the above considerations.

Theorem 2. Let there be given the union S of two similar particle systems $S_k$, $S_j$ as defined by Eqs. (2.1) – (2.4) of Sec. 2.1. Then the union system S is similar to its components $S_k$, $S_j$.

*Proof.* Let T be the temperature of $S_k$ and $S_j$. Then T is common to the systems S, $S_k$, $S_j$. The similarity of the system S to its components $S_k$ and $S_j$ follows from Theorem 1.

Corollary 1. From Theorems 1,2 it follows that

(2.8) $p_{k,i}(t)=p_{j,i}(t)$ for $i=1,...,I$, $t_0 \leq t < \infty$

or in the form

(2.8a) $n_{k,i}(t)/N_k = n_{j,i}(t)/N_j$

A simple algebra gives

(2.8b) $n_{k,i}(t)/n_{j,i}(t) = N_k/N_j$

and

(2.8c) $(n_{k,i}(t)+n_{j,i}(t))/N = n_{k,i}(t)/N_k = n_{j,i}(t)/N_j$, $N=N_k+N_j$

In [7] we studied the equivalent systems defined by common temperature, common particle density and other intensive quantities as the pressure, and by the same time evolution process. Let us denote the class of equivalent systems by $\{S_k^*\}$ $k=1,2...,K$, then $\{S_k^*\} \subset \{S_k\}$.

Remark 2. The similarity of the systems $S$, $S_1$, $S_2$ (one to the other) implies the equality of their occupation probabilities at each time $t_0 \leq t < \infty$.

Let us define the Shannon entropy by $S_e = -\sum_I p_i \ln p_i$ $i=1,2,..,I$, and the free energy by $a_e = E^* - k_B T S_e$ (see [7]).

Corollary 2. In virtue of Eq. (2.8c) we have

(2.9) $S_e(t) = S_e^j(t) = S_e^k(t)$, $a_e(t) = a_e^j(t) = a_e^k(t)$ for $t_0 \leq t < \infty$

This non-additivity of the entropy and the free energy results from the properties of the similar particle systems, Theorems 1,2 and Corollary 1.

The results concerning mixing of two isolated systems being in thermal equilibrium are summed up in Theorem 3, proven by Theorems 1,2 and Corollary 2.

Theorem 3. Let be given isolated particle systems $S_1$, $S_2$ having the same temperature $T_1 = T_2$.

Then

(1) Mixing of the systems $S_1$, $S_2$ forms at once the isolated system $S$ without any changes of the thermodynamic quantities. This mixing itself is not a thermodynamic process.

(2) The union system $S = S_1 + S_2$ is similar to its components.

(3) The uniformity of the system $S$ is expressed by Eqs. (2.8) – (2.8c).

Remark 3. We emphasize that the results of this section are independent of the particle number, particle type and of the volume of the systems. These points will be discussed in Sec. 3.

## 2.3. Mixing of the systems $S_1$ and $S_2$ having different initial temperature $T_1 \neq T_2$

We suppose, for example, $T_1 > T_2$.

To evoke the mixing process the partition XY (Fig. 1) is removed.

Our considerations are now based on the Clausius thermodynamic postulate and on the concept of thermal equilibrium [3,4].

Clausius postulate says that the heat is transferred from a hotter system to a colder one.

By thermal equilibrium we understand the state in which the energy exchange does not occur.

It is well known that the mixing process in the case of $T_1 \neq T_2$ is inherently connected with the energy exchange, the result of the particle interactions.

From the Clausius postulate it follows that the energy of the component $S_1$ decreases and the one of the $S_2$ component increases until thermal equilibrium, $T_1 = T_2$, is reached.

This indicates roughly that the particles belonging to the component $S_1$ pass from the higher energy levels to the lower ones. But in the component $S_2$ the "direction" is from the lower energy levels to the higher ones.

Note that the mixing considered in this section is a time-dependent process, which begins at the time $t_0$ and ceases at the time $t_e$, at which the thermal equilibrium is attained.

Hence the energies $E_1$ and $E_2$ are time-dependent quantities.

Because the system S is isolated the energy E is constant for $t \geq t_0$.

Eq. (2.3) becomes

(2.3a) $\quad E = E_1(t) + E_2(t) \quad$ for $\quad t_0 \leq t \leq t_e$

In the sequel it is shown that Eq.(2.3) is valid in the state of thermal equilibrium.

Clearly, the probabilities $p_{ki}$ and the mean energies $E_k^*$ are time-dependent for $t_0 \leq t \leq t_e$.

The mixing process considered here arises when the initial temperatures of the systems $S_1$ and $S_2$ are different, $T_1 \neq T_2$. This implies that the initial mean energies are of different values,

$E_1^*(t_0)$   $E_2^*(t_0)$. This process ceases when the equality $T_1=T_2$ ( $E_1^* = E_2^*$ ) is attained.

Now we proceed to investigate the thermal equilibrium state. From the Clausius postulate and the definition of the thermal equilibrium it follows that in thermal equilibrium the components $S_1$ and $S_2$ have equal temperature $T_1=T_2$ and the energies $E_1, E_2$ are constant. I.e. the temperature of the union system S is $T=T_1=T_2$ and its energy $E=E_1+E_2$.

From Theorems 1,2 it follows that the union system S and the components $S_1, S_2$ are similar one to the other. Of course Eqs. (2.8) - (2.8c) are valid.

Hence we conclude what follows. The mixing of two isolated systems $S_1, S_2$ of interacting particles having different temperature $T_1$ $T_2$, results in the union of similar systems $S=S_1+S_2$.

Now we shall determine the energies $E_1$, $E_2$ of the components $S_1$, $S_2$ and the temperature $T=T_1=T_2$ of the system S in thermal equilibrium state.

We recall

(2.10)     $E_1^* = E_2^*$ for $t \geq t_e$

Let us write Eq. (2.10) in the form

(2.10a)     $E_1/N_1 = E_2/N_2$  for $t \geq t_e$

A simple algebra gives

(2.10b) $E/N = E_1/N_1 = E_2/N_2$  for $t \geq t_e$

From equations $E=E_1+E_2$ and (2.10a) we obtain

(2.10c) $E_1 = N_1 E/N$, $E_2 = N_2 E/N$  for $t \geq t_e$

The temperature $T=T_1=T_2$ for $t \geq t_e$ is quite determined by the mean energy E/N of the whole system S.

The relations (2.10a) – (2.10c) allow us to calculate the quantities $E_1, E_2$ and $T=T_1=T_2$ of the systems in the thermal equilibrium state without difficult computations concerning the evolution of the thermalization process.

They express also the uniform distribution of the energy E and the temperature T over the whole system S.

Because Eqs. (2.8) - (2.8c) are valid, the uniform distribution of the particles over the energy levels $e_i$ i=1,...,I is given by Eq. (2.8c).

*Remark 4* The results of this section are independent of the extensive quantities $V_k$, $N_k$, $E_k$, k=1,2 and of the particle kind. Only the ratio $E_k/N_k$ plays an important role in the process considered here.

In order to sum up the results of this section we formulate below a theorem which is proven by the above considerations and equations. Note, the results of this section are quite general because the systems $S_1$, $S_2$ are arbitrary chosen.

Theorem 4. Let the systems $S_1$, $S_2$ and S be defined as in Sec. 2.1. Assume $T_1 \neq T_2$ at the time $t_0$, ($E_1^*(t_0) \neq E_2^*(t_0)$). Then:

1. In the case of $T_1 \neq T_2$ mixing is in essence the thermalization process, which ceases when thermal equilibrium is reached. I.e. when the equality $T_1=T_2$ ($E_1^*(t_e)=E_2^*(t_e)$) is attained.

2. This thermalization process results in the system $S=S_1+S_2$ similar to its components. This similarity arises simultaneously with the attainment of the thermal equilibrium. The uniformity of the union system S is expressed by the uniform distribution of the energy E within the whole system S, the temperature $T=T_1=T_2$ and the uniform distribution of the particles over the energy level $e_i$. See Eqs. (2.10) – (2.10c) and (2.8c).

3. The temperature $T=T_1=T_2$ and the energies $E_1,E_2$ of the systems $S_1,S_2$ in thermal equilibrium are determined by simple formulae which involve only the initial value $E_1(t_0)$ and $E_2(t_0)$.

Remark 7. When the system S reaches the thermal equilibrium, then the approach to the thermodynamic equilibrium begins (the Boltzmann distribution), with the initial conditions given by Eq. (2.8c).

The time evolution process has been investigated in [1,2].

**2.4. Discussion**

We have mentioned in Sec. 1 that the non-additive entropy appears in isolated particle systems composed of several such systems, S= $\sum_i S_i$ i 2. Theorems 3 and 4 show that the union S and its complements $S_1,S_2$ are similar each to the other. From Remarks 2-7 we have, the mixture S is independent of the particle numbers $N_i$, of the particle types and of the volumes $V_i$.

Therefore, we conclude that the non-additive entropy arises in every composed particle system S. For this reason we affirm: Entropy is a non-additive quantity.

The important point in our reasoning is that this statement is proven by the results of Secs. 2.2 and 2.3. The consequence following from this non-additivity is discussed in Sec. 3.

We note that the definition of Shannon entropy $S_e = -\sum_i p_i \ln p_i$ ($p_i$ are the occupation probabilities) is in agreement with the following Boltzmann's celebrated statement: The entropy can be expressed by the probabilities associated with the microscopic configuration of the system.

## 3. Remarks and conclusions

We have put in Sec. 2.1 the question, how mixing of two isolated systems $S_1, S_2$ containing interacting particles results in the uniform union system $S = S_1 + S_2$ defined by Eqs. (2.1) – (2.4). The answer to this question is given in Secs. 2.2 and 2.3.

The concept of similar particle systems along with the presentation of the system states by the occupation probabilities $p = (p_1, p_2, ... p_l)$ allow to understand the concept of non-additive entropy. The possible cases (i) $T_1 = T_2$ and (ii) $T_1 \neq T_2$ are here considered. Let us note the following points.

1. The relations between the temperatures $T_1, T_2$ govern the mixing process. From Theorems 2,3 and Corollary 1 it follows that in the case (i) mixing is not a thermodynamic process. The equality $T_1 = T_2$ establishes the sufficient and necessary conditions for the validity of Eqs. (2.8) – (2.8c) and (2.9).

   In the case (ii) mixing is a thermalization process, which ceases when the thermal equilibrium is reached. But both cases (i) and (ii) result in the uniform union system $S = S_1 + S_2$ similar to its components. The similarity and uniformity properties are characteristic for the particle systems being in thermal equilibrium.

2. In Secs. 2.2 and 2.3 it is shown that mixing of isolated particle systems is independent of the extensive quantities as volumes, energies, particle numbers and of the particle type. Only the mean energy, the ratio $E^* = E/N$ (= the temperature) plays an important role in this process. It is well-known that the time evolution process in isolated particle systems is dependent only on the temperature, see [7] and Appendix. The temperature determines uniquely the particle distribution in the thermodynamic equilibrium state. Thus the temperature plays a prominent role in the fundamental processes occurring in isolated systems of interacting particles.

3. The concept of non-additive entropy has been studied and examined through Secs. 2.1-2.4. The knowledge of similar particle systems along with the expression of the system state in terms of the occupation probabilities allow to understand the problem of non-additive entropy. In Secs. 2.2-2.4 it is shown:

   (i) The non-additive entropy rises in isolated, composed particle systems, $S = \sum_i S_i$.

   (ii) The properties of similar systems, given by Theorem 2 and Corollary 2, produce the non-additive entropy.

   (iii) The non-additive entropy is independent of the particle numbers $N_i$, the type of particles and the volumes $V_i$.

   We note that: (a) The non-additivity of the entropy is not in contradiction to other thermodynamic postulates and laws;

   (b) The entropy does not play any role in the course of the processes occurring in the particle systems.

   In virtue of the results obtained in Secs. 2.1-2.4 we may say what follows:

   The postulate of additive entropy should be replaced by the statement: The entropy is non-additive. From (a) and (b) we conclude that the classical thermodynamics with the concept of non-additive entropy is valid.

4. Mixing of the systems $S_1$, $S_2$ in both cases $T_1 = T_2$ and $T_1 \neq T_2$ results in an isolated, uniform system $S = S_1 + S_2$ (see Secs. 2.2 and 2.3). The time evolution to thermodynamic equilibrium in mixture S is an irreversible process, see [7].

5. The results of Secs. 2.2 and 2.3 may be generalized to arbitrary number of systems $S_k$, $k > 2$.

**Appendix. Time evolution process in isolated particle systems.**

In order to better understand Theorems 1,2 we recall some results of [7] concerning the approach to the thermodynamic equilibrium state.

Let S be an isolated particle system having the temperature T.

Let $t_0$ be the initial time (beginning of the time evolution process). Let the initial particle distribution (occupation probabilities) be $p_i(t_0)$, i=1,...I. The mean energy $E^* = \sum_i e_i p_i(t_0)$ is constant. The system state is defined by the finite sequence $p(t)=(p_1(t),p_2(t),...,p_I(t))$ $t_0 \leq t < \infty$, see Eq. (2.6). Thus the system state establishes a point in an I-dimensional Euclidean space. The temperature T determines the thermodynamic equilibrium state (Boltzmann distribution) $p_B=(p_{1,B},p_{2,B},...p_{I,B})$, where $p_{i,B}=e^{-\beta_i}/\sum_i e^{-\beta_i}$, $\beta=1/T$.

The equation governing the time evolution process is (see [7])

(A.1) $P_i(t)=\int_{t_0}^{t}[-\ln p_i(t)+a(t)+e_i b(t)]dt+p_i(t_o)$ i=1,...I  $t_0 \leq t < \infty$

The functions a(t) and b(t) are linear combinations of $\ln p_i(t)$ and are given explicitly in [1,2].

The system of Eq. (A.1) is denoted by (A.2).

The below items have been proven in [7].

1) The mapping (A.1) and (A.2) are one-one. The probabilities $p_i(t)$ are continuous, monotonic functions of the time t.

    There exists an inverse, one-one mapping to Eqs. (A.1) and (A.2).

    Thus we have the one-one correspondence t ---> $p_i(t)$ and $p_i(t)$ ---> t.

    That implies, the solution of the equation set (A.2) establishes a point p(t) in the I-dimensional Euclidean space, and $p_i(t)$ are its coordinates.

2) Let P be the range of the mapping (A.2). Therefore the coordinates of the point p(t) obey the two conservation laws, $\sum_i p_i(t)=1$, $\sum_i e_i p_i(t)=E^*$ ($E^*$ is a constant mean energy).

    Also the set P is an interval on the hyperline, intersection of two hyperplanes

    $\sum_i p_i(t)=1$, $\sum_i e_i p_i(t)=E^*$.

    The set P establishes the domain and the range of the mapping (A.2).

    D(A.2)=R(A.2).

3) To any time interval $<t_0,t_k>$ corresponds the interval $<p(t_o),p(t_k)>$, which belongs to the range of the equation set (A.2).

Let $t_k \rightarrow \infty$ as $k \rightarrow \infty$. Then $\sum_k <t_0,t_k> = <t_0, \infty)$ and $\sum_k <p(t_0),p(t_k)> = <p(t_0),p_B)$.

The approach to the thermodynamic equilibrium is asymptotic. The interval $<p(t_0),p_B)$ corresponds to the time interval $<t_0,\infty)$.

The time evolution process is closed under the thermodynamic equilibrium state. We say, the time evolution process takes place on the interval $<p(t_0),p_B>$.

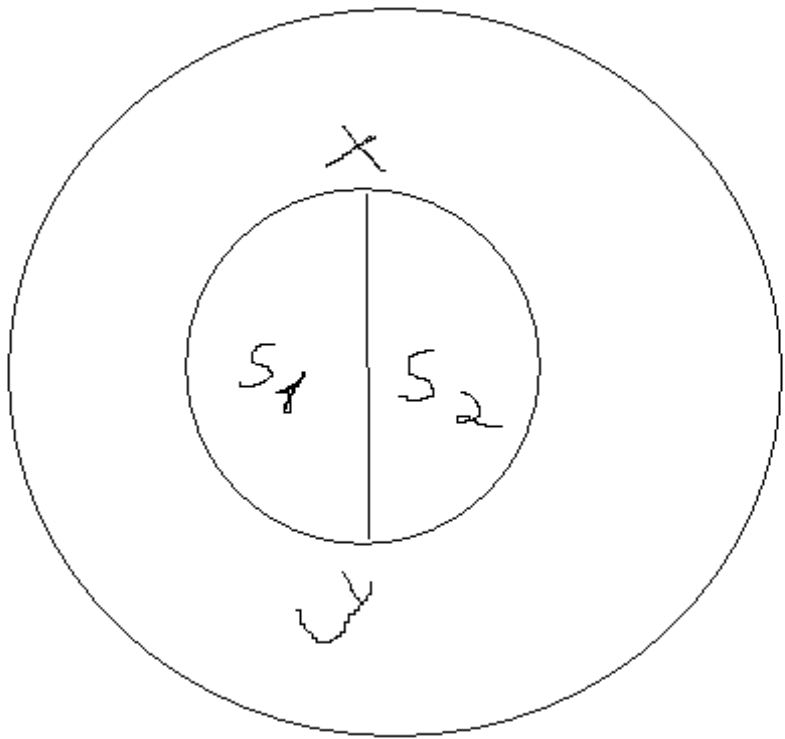

Figure 1. The union system $S=S_1+S_2$